\documentclass{emulateapj}
\usepackage{natbib}
\newcommand{\Msun}{{\rm M_{\odot}}}

\newcommand{\Zsun}{{\rm Z_{\odot}}}

\newcommand{\kpc}{\, {\rm kpc}}
\newcommand{\pc}{\, {\rm pc}}
\newcommand{\kmps}{\, {\rm km \, s^{-1}}}

\newcommand{\HII}{\mbox{H\,{\sc ii}}}

\slugcomment{Submitted to ApJL}

\shorttitle{Discovery of New Dwarf Galaxy near NGC~6503}
\shortauthors{Koda et al.}

\begin{document}

\title{Discovery of New Dwarf Galaxy near The Isolated Spiral Galaxy NGC~6503}

\author{Jin Koda\altaffilmark{1},
Masafumi Yagi\altaffilmark{2},
Yutaka Komiyama\altaffilmark{2},
Samuel Boissier\altaffilmark{3},
Alessandro Boselli\altaffilmark{3},
Alexandre  Y. K. Bouquin\altaffilmark{4},
Jennifer Donovan Meyer\altaffilmark{5},
Armando Gil de Paz\altaffilmark{4},
Masatoshi Imanishi\altaffilmark{6},
Barry F. Madore\altaffilmark{7},
David A. Thilker\altaffilmark{8}}

\email{jin.koda@stonybrook.edu}

\altaffiltext{1}{Department of Physics and Astronomy, Stony Brook University, Stony Brook, NY 11794-3800}
\altaffiltext{2}{Optical and Infrared Astronomy Division, National Astronomical Observatory of Japan, 2-21-1 Osawa, Mitaka, Tokyo, 181-8588, Japan}
\altaffiltext{3}{Aix Marseille Universit\'e, CNRS, LAM (Laboratoire d'Astrophysique de Marseille) UMR 7326, 13388, Marseille, France}
\altaffiltext{4}{Departamento de Astrof\'{\i}sica y CC. de la Atm\'{o}sfera, Universidad Complutense de Madrid, Avda. de la Complutense s/n, Madrid E-28040, Spain}
\altaffiltext{5}{National Radio Astronomy Observatory, 520 Edgemont Road, Charlottesville, VA 22903, USA}
\altaffiltext{6}{Subaru Telescope, 650 North A'ohoku Place, Hilo, HI 96720}
\altaffiltext{7}{Carnegie Institution of Washington, 813 Santa Barbara Street, Pasadena, CA 91101}
\altaffiltext{8}{Center for Astrophysical Sciences, The Johns Hopkins University, 3400 N. Charles Street, Baltimore, MD 21218}

\begin{abstract}
We report the discovery of a new dwarf galaxy (NGC~6503-d1) during the Subaru extended ultraviolet (XUV) disk survey.
It is a likely companion of the spiral galaxy NGC~6503.
The resolved images, in the $B$, $V$, $R$, $i$, and H$\alpha$ bands, show an irregular appearance due to
bright stars with underlying, smooth and unresolved stellar emission.
It is classified as the transition type (dIrr/dSph) between the dwarf irregular (dIrr) and dwarf spheroidal (dSph) types.
Its structural properties are similar to those of the dwarfs in the Local Group, with an absolute
magnitude $M_{\rm V}\sim  -10.5$, half-light radius $r_{\rm e}\sim 400\pc$, and central surface brightness $\mu_{0, \rm V}\sim 25.2$.
Despite the low stellar surface brightness environment, one $\HII$ region was detected,
though its H$\alpha$ luminosity is low, indicating an absence of any appreciable O-stars at the current epoch.
The presence of multiple stellar populations is indicated by the color-magnitude diagram of $\sim 300$ bright resolved stars
and the total colors of the dwarf, with the majority of its total stellar mass $\sim 4\times 10^6 \Msun$ in
an old stellar population.
\end{abstract}


\section{Introduction}
Dwarf satellite galaxies provide vital guidance for understanding star formation (SF) and galaxy evolution,
and their discovery and structural analysis are of broad interest.
They often have extremely low surface brightness (LSB) and exhibit rare instances of SF
in the low-density environments.
They are potential triggers of SF
in their more prominent parent galaxies.
The observed statistics and internal structures of dwarf satellites in the Local Group
are challenging the $\Lambda$CDM scenario \citep[][]{Klypin:1999lr, Moore:1999fk, Boylan-Kolchin:2012bk}.
Systematic searches for dwarf satellites around other galaxies, including a more rigorous one in the Local Group,
are being undertaken \citep[e.g., ][ Donovan Meyer et al. submitted]{McConnachie:2012fk, Monachesi:2014zr, Merritt:2014rt, Sand:2014yq, Tollerud:2014fk, McQuinn:2015fk}.

In this Letter, we report the discovery of a new dwarf galaxy NGC~6503-d1, approximately $17\arcmin$
away from the spiral galaxy NGC~6503.
NGC~6503 is comparatively isolated at the edge of the Local Void \citep{Karachentsev:2003uq, Greisen:2009lr}.
It is an SA(s)cd type spiral galaxy at a distance of $d$=$5.27\pm0.53$ Mpc \citep[$1\arcsec=25.5$ pc; ][]{Karachentsev:2003uq}
with a heliocentric velocity of $v_{\rm h}=25\pm1 \kmps$ \citep{Epinat:2008fk}.
NGC~6503 has a major axis of $D_{25} \sim 10.9\kpc$ ($7.1\arcmin$)
and an absolute magnitude of $M_{\rm V}=-19.07$ \citep[calculated from ][]{de-Vaucouleurs:1991lr}.
Assuming a mass-to-light ratio 
$\gamma_{\rm V} \sim 1.2$ \cite[see ][]{McGaugh:2014qy},
NGC~6503's stellar mass is $\sim 4\times 10^9 \Msun$,
and its virial radius is approximately $150\kpc$ \cite[see ][]{Kravtsov:2013uq}.
We present the basic properties, stellar populations, and recent SF activity of the new dwarf galaxy
NGC~6503-d1.
Uncertainty still remains as to the distance of NGC~6503-d1, but based on its proximity and
other supporting evidence the dwarf is most likely associated with NGC~6503.

NGC~6503 shows very extended ultraviolet emission far beyond the edge of its optical disk \citep{Thilker:2007dp}.
Such extended ultraviolet (XUV) disks indicate SF at a very low density \citep{Koda:2012ab} and
are quite common \citep[residing in $\sim$30\% of nearby galaxies; ][]{Thilker:2007dp}.
However, the trigger of the XUV activity still remains unknown
\citep[e.g.,][]{Espada:2011fj}. NGC~6503-d1 is the first dwarf satellite discovered during the early observations of
the Subaru-XUV survey, a deep optical imaging survey of nearby XUV disks using the Subaru telescope.

\section{Observations and Data Reduction}
The dwarf galaxy was discovered in deep imaging of the spiral galaxy NGC~6503
using the the Subaru Prime Focus Camera \citep[Suprime-Cam; ][]{Miyazaki:2002fk}
on the Subaru telescope.
Suprime-Cam has a mosaic of ten $2048 \times 4096$ CCDs,
covering a wide field of $34\arcmin \times 27\arcmin$ with a pixel scale of $0.20\arcsec$.
The observations were performed in June and July 2013 and supplemented with archival data taken in July 2012.
The field of NGC~6503 was observed in the $B$, $V$, $R$, $i$, and NA656 (H$\alpha$) bands,
among which the images in $V$ and $i$ were deeper.
The typical seeing was 0.8$\arcsec$.

The data were reduced in a standard way, including an overscan correction, masking of the shadow
of the autoguider probe, flat-fielding, and distortion correction before co-adding all exposure frames \citep{Yagi:2002lr}.
Photometric calibrations were performed using a catalog
from the Panoramic Survey Telescope \& Rapid Response System 1
\citep[Pan-STARRS~1; ][]{Magnier:2013lr, Schlafly:2012fk, Tonry:2012qy}.
We followed the procedure described in \cite{Yagi:2013fk} and converted
from the Pan-STARRS~1 to the Subaru band system using stellar photometry.
Typical photometric errors including both statistical and systematic effects are $\lesssim 0.1$ mag.

We also obtained archival FUV and NUV images from the Galaxy Evolution Explorer (GALEX).
We averaged two available data sets from the GALEX Nearby Galaxies Survey \cite[NGS; ][]{Gil-de-Paz:2007lj}
and from the General Investigator Program (GI1 Program 097).

\section{Discovery and Classification}

\begin{figure*}
\epsscale{1.10}
\plotone{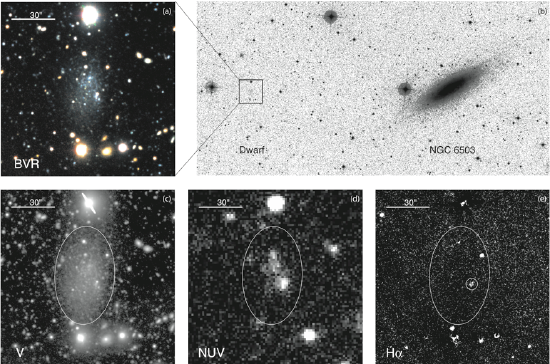}
\caption{
Subaru images of $2\arcmin \times 2\arcmin$ region around the dwarf,
except at top-right, which is a $\sim 30.2\arcmin \times 12.7\arcmin$ region
from the Digitized Sky Survey (DSS).
North is up, and east is left.
The photometric aperture is an ellipse with a major axis of
$\sim 68\arcsec$, axis ratio of 0.61, and position angle of $-2.6\deg$ from north to east.
(a) Pseudo color image with Subaru $B$, $V$, and $R$.
(b) DSS image of NGC~6503 and the new dwarf galaxy.
(c) Subaru $V$ band image in log scale.
(d) GALEX NUV image.
(e) Subaru H$\alpha$ image (net H$\alpha$).
The small circle encloses the identified $\HII$ region.
\label{fig:images}}
\end{figure*}

Figure \ref{fig:images} shows the new dwarf galaxy NGC~6503-d1.
The $BVR$ color image (panel a) shows a lopsided distribution of resolved bright stars in
the northern part of the dwarf, apparently indicating an irregular morphology.
However, we also found a smooth, extended, and symmetric stellar distribution in logarithmic scale
(panel c; $V$ band example), which is perhaps a more accurate
representation of the stellar mass distribution.
Panel (e) shows the continuum-subtracted H$\alpha$ image.
H$\alpha$ emission is detected at one spot within the dwarf.
The presence of the resolved bright stars and the $\HII$ region indicate that
this dwarf hosts recent/current SF and, by implication, some gas,
though no HI gas was detected at the $3\sigma$ upper limit of $<10^6\Msun$ \citep{Greisen:2009lr}.

We classify this as a transition-type dSph/dIrr, based on traditional definitions
\citep{Tolstoy:2009lr, McConnachie:2012fk}, the type between dwarf irregular (dIrr)
and dwarf spheroidal (dSph).
We note, however, that there is a discussion on whether dIrrs and dSphs are
physically distinct, the only difference being their current/recent SF which outshines
the other populations and gives the impression of an irregular morphology
\citep[e.g., ][]{Weisz:2011fk}.

\section{Properties}

NGC~6503-d1 is very likely a satellite galaxy of NGC~6503.
Their projected separation $\sim 17\arcmin$ is well inside NGC~6503's
virial radius $\sim 100\arcmin$ ($\sim 150\kpc$), suggesting a high probability of their physics
association ($>80\%$) based on $N$-body simulations \citep[e.g., ][]{Tollerud:2011lr}.
The H$\alpha$ detection also supports its proximity to NGC~6503,
though the heliocentric velocity of NGC~6503 ($v_{\rm h}=25\pm1 \kmps$) is very close to
MW's and it is difficult to distinguish between NGC~6503 and the MW.
Although uncertainty still remains, we assume that the dwarf is a companion of NGC~6503
in what follows. We will discuss a few pieces of supporting evidence in
addition to the proximity, including its size and magnitude (distance-dependent parameters)
compared to other dwarf galaxies (\S \ref{sec:stat})
and a likely detection of the tip of red giant branch (TRGB) star population (\S \ref{sec:resolved}).
Each of these alone is insufficient to prove that the dwarf is a satellite
of NGC~6503.The combination, however, supports our working assumption,
that $d=5.25$ Mpc is the distance to NGC~6503-d1.

\begin{deluxetable}{lc}
\tablecolumns{2}
\tablecaption{Properties of NGC~6503-\lowercase{d}1}
\tablehead{ \colhead{Parameter} & \colhead{Value} }
\startdata
\cutinhead{Distance-independent Properties}
R. A. (J2000) & 17:52:48.4  \\
DEC (J2000) &  +70:08:14.4\\
Galactic longitude $l$ (degree) & 100.5403\\
Galactic latitude $b$ (degree) & 30.3563 \\ 
$V_{\rm h}$ ($\kmps$) & $\sim 0$ \\
FUV (AB mag) & 21.1\\
NUV (AB mag) & 20.4\\
B (AB mag) & 18.5 \\
V (AB mag) & 18.1 \\
R (AB mag) & 17.8 \\
i (AB mag)  & 17.6 \\
Sersic index $n$ & 0.54 \\
$\mu_{\rm 0, V}$ ($\rm AB\, mag\, arcsec^{-2}$) & 25.2 \\
$r_{\rm e}$ ($\arcsec$) & 15.6 \\
Axis ratio & 0.61 \\
P. A. (deg, N to E) & -2.6 \\
\cutinhead{Distance-dependent Properties}
$m$-$M$ (mag) & 28.61$\pm$0.23\phn \\
$d$ (Mpc) & 5.27$\pm$0.53\phn \\
$M_{FUV}$ (AB mag) & -7.6 \\
$M_{NUV}$ (AB mag) & -8.2 \\
$M_B$ (AB mag) & -10.1 \\
$M_V$ (AB mag) & -10.5 \\
$M_R$ (AB mag) & -10.8 \\
$M_i$ (AB mag) & -11.0 \\
$r_{\rm e}$ ($\pc$) & 400 \\
$F_{H\alpha}$ ({$\rm erg\, s^{-1}$}) & $8.2\times 10^{35}$\\
Stellar mass ($\Msun$) & $\sim 4 \times 10^6$  \\
HI gas mass ($\Msun$) & $<10^6$
\enddata
\label{tab:prop}
\tablenotetext{}{
All magnitudes are corrected for Galactic extinction.
Typical errors due to photometric calibration and background
contamination are $\sim 0.14$ mag.}
\end{deluxetable}

\subsection{Internal Structure}

The GALFIT package \citep{Peng:2002fk} was adopted to measure structural parameters
of the dwarf. The fit was made to the $V$ band image with a Sersic profile
\citep{Sersic:1968uq} and fixed sky background after removing some surrounding stars.
We obtained a Sersic index of $n=0.54$ with $\chi^2_{\nu}=2.8$ ($\nu=90494$).
This result was consistent with that using the $i$ band image.
As noted earlier, some bright stars within the dwarf outshine the underlying smooth stellar distribution.
After aggressively masking these we obtained $n=0.63$ ($\chi^2_{\nu}=0.63$)
though it is debatable whether the member stars of the dwarf should be masked out.
Table \ref{tab:prop} gives properties of NGC~6503-d1, including its central coordinates, axis ratio,
position angle (P. A.), and effective radius/half-light radius $r_{\rm e}$, as well as other global parameters.
The central surface brightness $\mu_{\rm V}$ was derived from the value at the central
pixel of the fitted model, and the result after Galactic extinction correction,
$\mu_{\rm V}=25.18 \,\rm mag\,arcsec^{-2}$, is typical for dwarf galaxies \citep{McConnachie:2012fk}.

Apparent magnitudes (AB magnitude) were calculated within an elliptical aperture (Figure \ref{fig:images}).
The shape of the aperture was derived from GALFIT (i.e., an axis ratio and P. A.) with the major axis diameter scaled to $34\arcsec$
so as to enclose the emission of the dwarf.
The fluxes of foreground/background contamination were estimated by placing the apertures in
the surrounding region. We subtracted their averages from the fluxes within
the dwarf aperture and adopted their typical scatters of 0.1 mag as errors in photometry.
The apparent magnitudes are listed in Table \ref{tab:prop}.
We adopted a Galactic extinction of $A_{\rm V}=0.106$ \citep{Schlegel:1998zk, Schlafly:2011qy}
and calculated extinctions in UV and NA656 with ($A_{\rm FUV}$, $A_{\rm NUV}$) = (2.58, 2.83)$A_{\rm V}$
for the Milky Way \citep{Koda:2012ab} and $A_{\rm NA656} = A_{\rm R}$.
These aperture magnitudes were consistent with the ones from GALFIT (e.g., 18.2 mag in $V$).

\subsection{Statistical Properties}\label{sec:stat}

\begin{figure*}
\epsscale{0.8}
\plotone{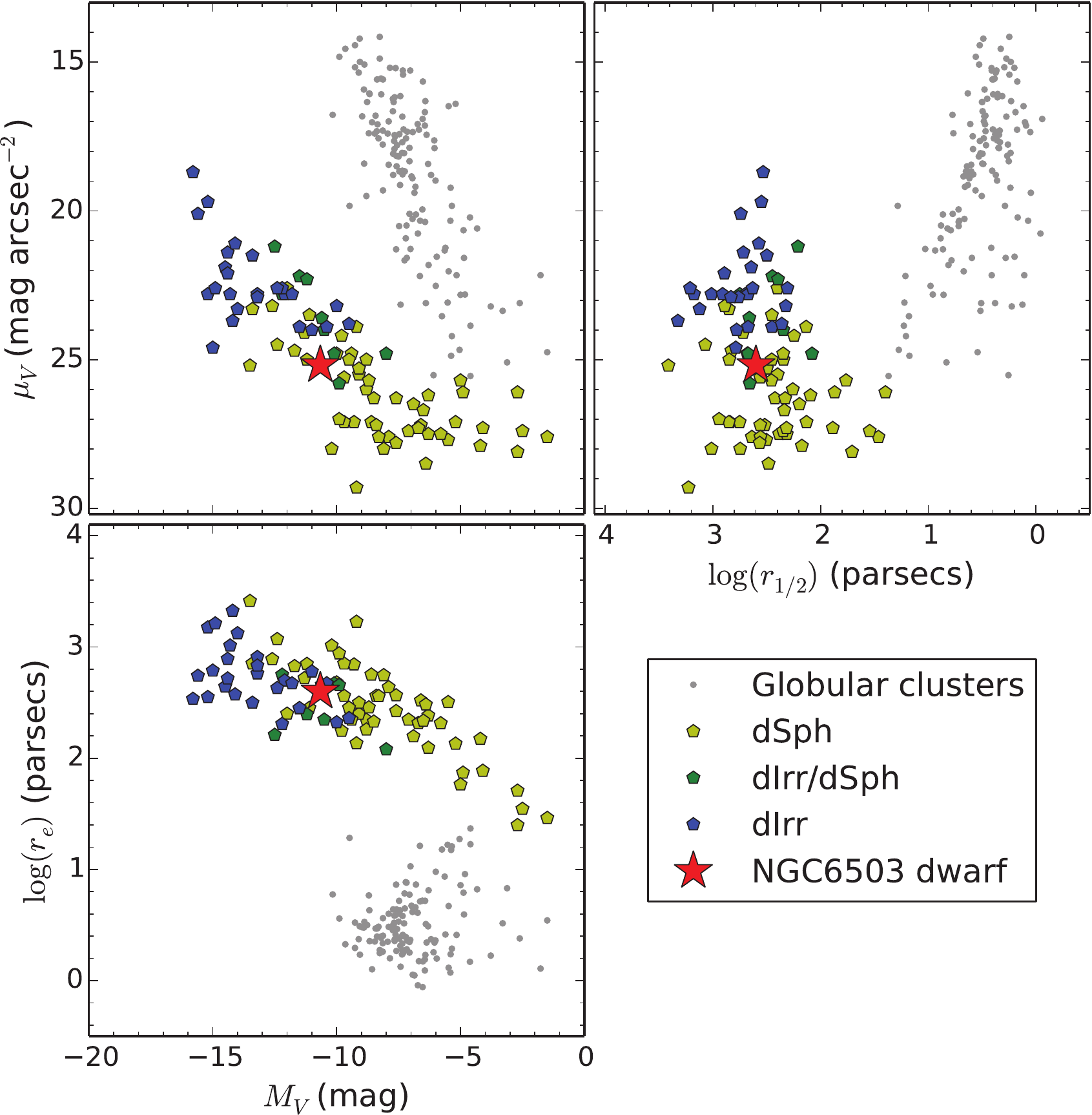}
\caption{
Correlations between central surface brightness $\mu_{\rm V}$, half-light radius $r_{\rm e}$, and absolute $V$ magnitude $M_{\rm V}$.
The red star is the new dwarf galaxy. Comparison data are dwarf spheroidal (dSph), dwarf irregular (dIrr), and transition (dIrr/dSph) galaxies
in the Local Group compiled by \cite{McConnachie:2012fk} and Galactic globular clusters from \cite{Harris:1996fk}.
We adopted the distance of NGC~6503 for NGC~6503-d1.
\label{fig:corr}}
\end{figure*}

If it is a satellite of NGC~6503, NGC~6503-d1 lies within the sequence of dwarf galaxies in the Local Group,
in surface brightness $\mu_{\rm V}$, absolute magnitude $M_{\rm V}$,
and half-light radius $r_{\rm e}$.
Figure \ref{fig:corr} shows their correlations with Local Group dwarfs \citep{McConnachie:2012fk}
and Galactic globular clusters \citep{Harris:1996fk} as references. The new dwarf is distinct from
the globular clusters and lies midway between dIrrs (blue) and dSphs (yellow) and among the dIrr/dSph transition types (green).
This location is consistent with the visual appearance of the dwarf,
having apparent irregular morphology (due to recent SF) with the more symmetric morphology of
the underlying stellar population.

The Sersic profile of $n=0.54$ is also within the range typical of dwarf galaxies of similar brightness.
Radial profiles of dwarfs are often fit with an exponential profile
\citep[an $n=1$ profile; e. g., ][]{Ibata:2007fj, Richardson:2011uq, Crnojevic:2014qy},
though $n=0.3$-$0.7$ has been reported for dwarf galaxies around M101 \citep{Merritt:2014rt}
and an isolated dwarf galaxy \citep{Monachesi:2014zr}.
\citet{Geha:2006fj} showed a spread of $n\sim0.5$-$1.5$, albeit among brighter dwarfs ($\lesssim -14$ mag).
The structural properties of the new dwarf are, therefore,
typical of faint dwarf galaxies in the Local Group and in some other nearby systems.

\subsection{Resolved Stars}\label{sec:resolved}

\begin{figure*}
\epsscale{0.95}
\plotone{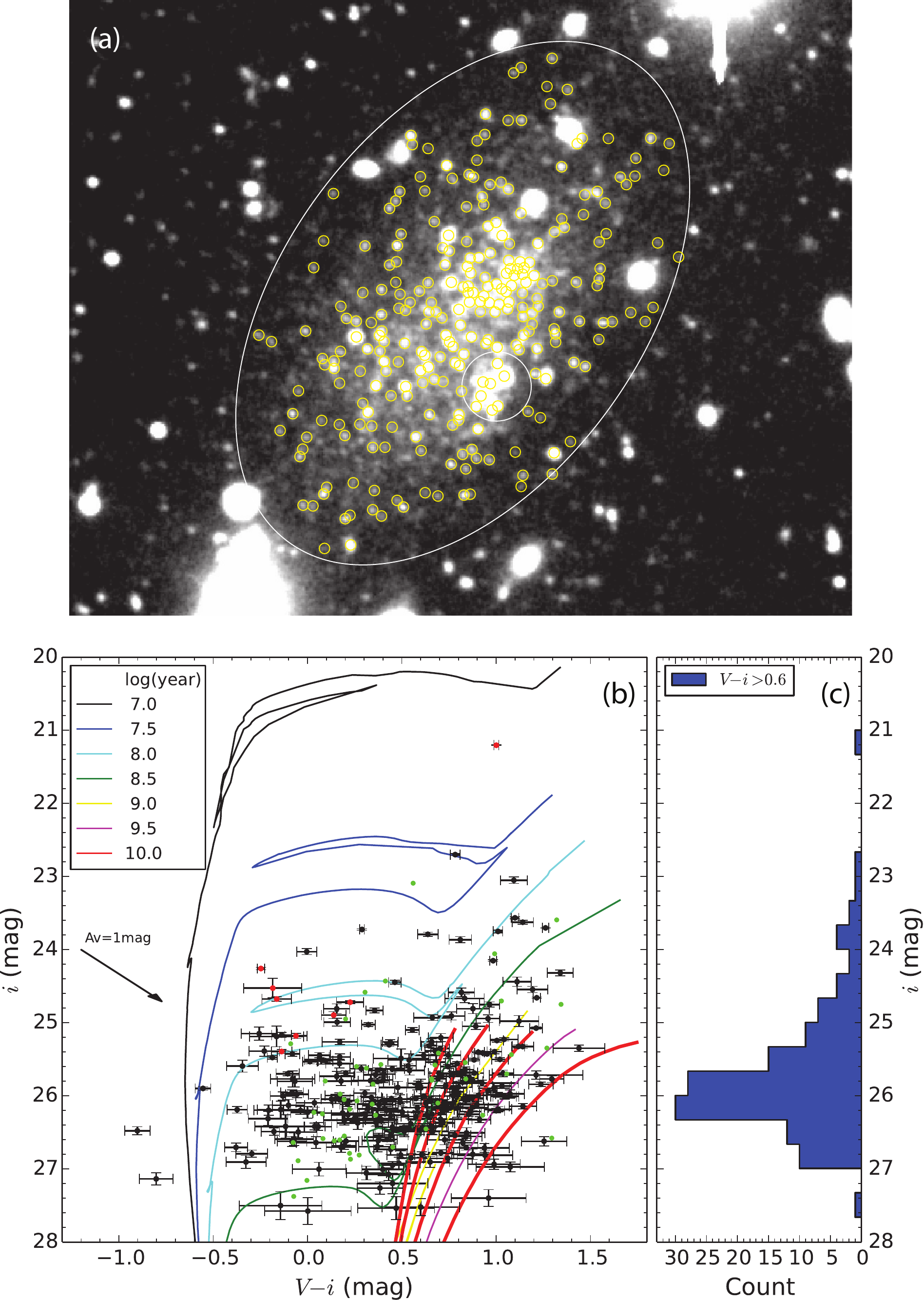}
\caption{Analysis of resolved stars.
(a) Locations of resolved stars (yellow circles) in $i$ band image.
The image orientation is not rotated from the original to maintain the high spatial resolution.
The while ellipse and circle (HII region) are the same as the ones in Figure \ref{fig:images}.
(b) Color-magnitude diagram (CMD) with $V$ and $i$ bands.
The stellar isochrones are from
the Padova evolutionary tracks (CMD 2.7) and of $\log \rm (Z/Z_{\odot})=-0.5$
at $\log(\rm Age)=$7.0, 7.5, 8.0, 8.5, 9.0, 9.5, and 10.0.
The red isochrones are of $\log(\rm Age)=$10.0
for $\log \rm (Z/Z_{\odot})=$-2.0, -1.5, -1.0, and -0.5 from left to right, respectively.
We applied the distance modulus of NGC~6503 ($m-M=28.62\pm0.23$).
Stars within the HII region aperture (white circle in panel (a)) are marked red.
For comparison, the green dots are objects in a control field of the same size
about $2\arcmin$ away from NGC~6503-d1 (at the same distance from NGC~6503);
these background objects are populated much more sparsely in the CMD (this is so, even though
the difference in confusion limit is not corrected).
(c) Histogram of apparent $i$ band magnitude for all identified stars of $V-i>0.6$.
The increment is 0.33 mag. From the bright to faint, the number starts increasing
between the bins of $m_i=25.17$ and $25.50$.
The AB magnitude of TRGB in Subaru $i$ is approximately $-3.53$ (absolute) and $25.09\pm0.23$ (apparent; see the text).
 \label{fig:cmd}}
\end{figure*}

The high spatial resolution of the Subaru images ($\sim 0.8\arcsec$) permits identification and
photometry of individual bright stars within NGC~6503-d1.
We made an initial catalog of 312 potential detections using the DAOPHOT PSF
photometry software \citep{Stetson:1987qy}. We then manually rejected apparent misidentifications
(e.g., background galaxies and tails between adjacent bright stars), leaving 303 stars,
which are identified in Figure \ref{fig:cmd}a.
We performed photometry in $V$ and $i$ bands, made a color-magnitude diagram (CMD; {\it middle})
and a histogram of apparent $i$ magnitude for stars with $V-i>0.6$ ({\it right}).

Theoretical isochrones are superimposed on the CMD assuming a distance of NGC~6503 ($m-M=28.62\pm0.23$)
and were taken from the Padova and Trieste stellar evolutionary tracks
\citep{Bressan:2012uq, Chen:2014qy, Tang:2014fj}\footnote{http://stev.oapd.inaf.it/cgi-bin/cmd}.
The red isochrones are at an age of 10 Gyr for metallicities of $\log (\rm Z/Z_{\odot})=$ -2.0, -1.5, -1.0, and -0.5
from the left to right.
The others are for $\log \rm (Z/Z_{\odot})=$-0.5 and at $\log ({\rm Age [year]})$=7 to 10 (see legend).
The bright stars that we identified are around the areas of the asymptotic giant branch (AGB), horizontal branch (HB),
and red giant branch (RGB) sequences. Therefore, the dwarf hosts multiple stellar populations,
from young ($\sim$100 Myr, cyan isochrone) to old ($\gtrsim$8-10 Gyr, magenta and red).
Many stars in the recently formed population must have escaped their natal dust given the agreement
between unreddened isochrones and our photometry, especially regarding the blue limit of the He-core-burning population.
The RGB sequence depends on stellar metallicity \citep[red lines; see also ][]{Lee:1993lr}.
The spread on the red side may indicate a variety of metallicity in the underlying stellar population.

The tip of RGB (TRGB) stars are often used as a distance indicator, as their luminosities are approximately
constant near the $i$ band wavelength \citep{Lee:1993lr}. By averaging the four red isochrones in Figure \ref{fig:cmd}b
we determined a TRGB magnitude of $M_{i}=-$3.53 AB mag in the Subaru $i$ band.
If the dwarf is a companion of NGC~6503,
the corresponding apparent magnitude is $m_{i}=25.09$ with an error of at least $\pm0.23$ carried over from the distance modulus.
Within the errors, this is roughly where we see the rise in the histogram (Figure \ref{fig:cmd}c; $V-i>0.6$ for TRGB color),
though the sample is small. This supports the assumption that the dwarf is a companion of NGC~6503.

\subsection{Star Formation History}\label{sec:sfh}

The central and average  $V$ band surface brightnesses are
25.18 and 26.31$\,\rm mag\, arcsec^{-2}$, respectively, after correcting for the Galactic extinction.
Any SF there is a rare incident in a very LSB regime.
Conventionally, galaxies sizes have been defined by the 25$\,\rm mag\, arcsec^{-2}$ isophote level.
The distribution of resolved stars in the CMD
suggests multiple stellar populations, indicating that SF has persisted for a long period of time.
Figure \ref{fig:images}e shows the detection of one $\HII$ region (small circle) within the dwarf
at $6\sigma$ significance.
The $\HII$ region is virtually co-incident with the brightest spot in NUV (GALEX J175247.4+700811; Figure \ref{fig:images}d),
indicating current SF in this LSB galaxy.
[The brightest H$\alpha$ spot at NW without NUV emission is unlikely to be in the dwarf, since it would be too bright.]

The SF activity of this $\HII$ region is low, which is inferred from its faint H$\alpha$ luminosity
of $L_{\rm H_{\alpha}}=8.2\times 10^{35} \,\rm erg\,s^{-1}$.
This was measured within an $8\arcsec$-diameter aperture centered at
($\alpha_{\rm J2000}$, $\delta_{\rm J2000}$) = (17:52:47.1,+70:08:09.8) with correction for
Galactic extinction.
The uncertainty is about a factor of 2 due to a sensitivity to continuum subtraction
(The fraction of H$\alpha$ over total emission in the NA656 band is $\sim$5\% in the aperture).
The contamination of [NII] $\lambda$ 6548, 6583 emission is
likely negligible \citep[see ][]{Kennicutt:2008kx}.
This $L_{\rm H_{\alpha}}$ is on the order of that expected around a single B0V
type star \citep[$\sim 7.0\times10^{35} \,\rm erg\,s^{-1}$ under case B recombination
with an electron temperature of $10^4$ K; ][]{Sternberg:2003yb, Osterbrock:2006pd, Koda:2012ab}.
This is consistent with observations of the Orion Nebula in the Milky Way,
which is about 10 times brighter in $L_{\rm H_{\alpha}}$ with only a single O5.5V type star
as a dominant ionizing source. The second brightest stars in Orion \citep[one O9.5V, two B0.5V, and one B3; ][]{Reipurth:2008yq}
together emit $\sim 10$-$20$ times fewer Lyman continuum photons \citep[e.g., ][]{Sternberg:2003yb}.
Therefore, the dwarf galaxy is unlikely to be hosting any appreciable O-star population at the current time,
assuming that the leakage of ionizing photons is not dominant
($\gtrsim 90\%$ of the Lyman continuum photons need to leak away without ionizing the surrounding gas to hide an O-star).
Of course, it could have had O-star(s) a short time ago as their lifetimes are short  ($\sim 10$ Myr).
The stars around this HII region are AGB and blue loop stars (Figure \ref{fig:cmd}).
One, or some, of these may be the ionizing source if an internal extinction
is locally large ($A_{\rm V}\sim1$).

The observed colors of the dwarf (Table \ref{tab:prop}) also indicate multiple epochs of SF.
Rigorous analyses of SF history \cite[e.g., ][]{Weisz:2011fk, Weisz:2012lr} are beyond our scope,
and a dwarf galaxy of small mass requires stochasticity in SF to be taken into account
\citep[e.g., ][]{da-Silva:2012lr}.
Instead, we adopt a simplistic approach to illustrate that 
at least two stellar populations, old and young, are necessary to account for the observed colors,
and that most of the stellar mass is old.
We adopted Starburst99 \citep{Leitherer:1999qh}, a Kroupa initial mass function,
a metallicity of $Z=0.4\Zsun$, and an internal extinction $A_{\rm V}=0.1$ for the integrated colors of the dwarf.
These choices were made arbitrarily for demonstration;
for example, the metallicity may have a range from a low $\log(Z/\Zsun)$ up to $>-0.5$
according to Figure \ref{fig:cmd} (the data with respect to the RGB isochrones).

\begin{figure}
\epsscale{1.1}
\plotone{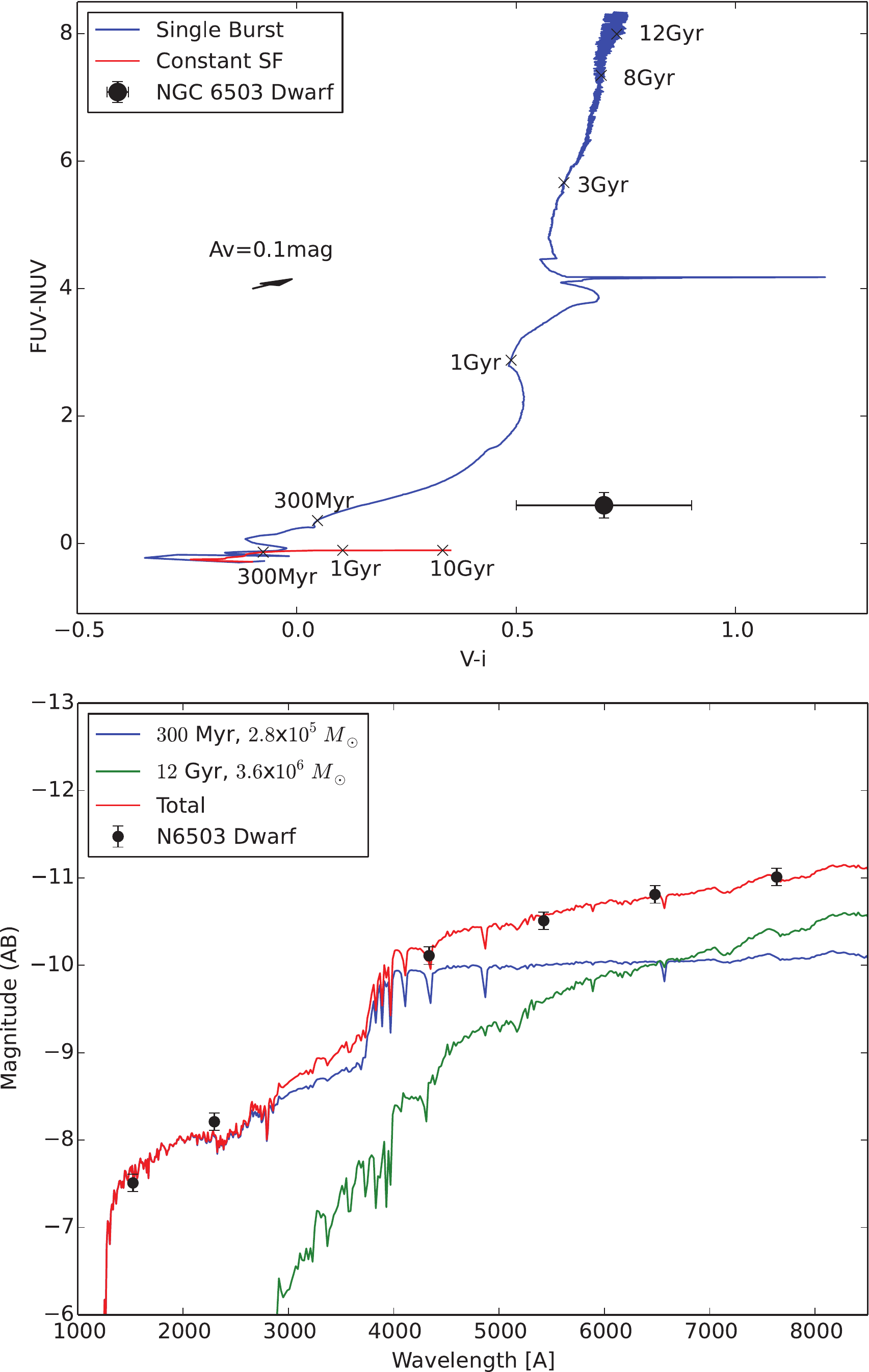}
\caption{
Comparisons of photometric evolution models (Starburst99) and the data for the dwarf.
A metallicity of $0.4\rm Z_{\odot}$ and a Kroupa initial mass function are adopted.
(a) FUV-NUV vs V-i plot.
A single burst model (blue) and constant SF model (red) are plotted.
The arrow shows an internal extinction of $A_{\rm V}=0.1$ mag
using the extinction curve of LMC \citep{Pei:1992fh}, which is not applied to the data
nor models.
(b) Spectra from the single burst model at ages = 300 Myr (blue) and 12 Gyr (green),
and the summation of the two (red).
A correction for internal extinction ($A_{\rm V}=0.1$) was applied.
The zero-points of the spectra (i.e., stellar masses) were adjusted by eye to approximately
fit the observed FUV, NUV, $B$, $V$, $R$, and $i$ magnitudes of the dwarf.
\label{fig:spec}}
\end{figure}

Figure \ref{fig:spec}a shows that neither a model of single starburst at any age
nor continuous SF alone reproduces
the FUV-NUV and V-i colors simultaneously unless  $A_{\rm V}$ is unusually high.
Figure \ref{fig:spec}b shows an example of stellar populations at the young and old ages
of 300 Myr and 12 Gyr.
The young population (blue) is relatively bright in FUV and NUV
and shows a flat spectrum in the redder, optical range ($B$, $V$, $R$, and $i$).
The old population (green) contributes negligibly in the FUV and NUV and shows a rising spectrum in the optical bands.
The dwarf is relatively bright in FUV and NUV (requiring a young stellar population)
and shows a rising trend in the optical bands (old stellar population).
The masses of stellar populations determine the zero point in magnitudes:
the old component carries the most mass ($\sim  3.6 \times 10^6 \Msun$; insensitive
to the actual age if $\gtrsim8$ Gyr) as compared to the young one ($\sim  2.8 \times 10^5 \Msun$).
The total stellar mass within the dwarf is approximately $4\times10^6\Msun$.

\section{Summary}
We have discovered the new dwarf galaxy of the dIrr/Sph transition type, NGC~6503-d1,
which is likely a physical companion of the spiral galaxy NGC~6503.
Its structural properties are similar to those of the Local Group dwarfs.
Integrated colors and resolved stars indicate a complex SF history
not described by a single burst of any particular age nor by a simple continuous SF,
which is also similar to the other known dwarfs \citep{Weisz:2011fk, Weisz:2012lr}.
The majority of the total stellar mass $\sim4\times10^6\Msun$ is in an old stellar population.
NGC~6503-d1 is the only companion galaxy of this magnitude near NGC~6503 within
our field of view, about 1/6 of the virial radius.
It is possible, and likely, that more dwarfs can be found in a future survey.

\acknowledgments
This research used data collected with the Subaru Telescope operated by the National Astronomical
Observatory of Japan and with the GALEX operated for NASA
by the Caltech under NASA contract NAS5-98034,
and used the NASA/IPAC Extragalactic Database (NED) operated by
the Jet Propulsion Laboratory, Caltech, under contract with NASA.
We thank an anonymous referee for constructive comments,
Fumiaki Nakata for help in observations, Eric Greisen for HI data, and
the Pan-STARRS1 Surveys collaboration for the photometric catalog.
This research is supported by NASA  through grant NNX14AF74G.
JK also acknowledges support from the NSF (AST-1211680).




\end{document}